\documentclass[prb,twocolumn,showpacs,amsmath,amssymb,superscriptaddress]{revtex4}
\usepackage{graphicx}
\usepackage{enumerate}
\newcommand \beq{\begin{eqnarray}}
\newcommand \eeq{\end{eqnarray}}

\newcommand{\bfr}{\mathbf{r}}
\newcommand{\bfk}{\mathbf{k}}

\newcommand{\bfs}{\mathbf{s}}

\newcommand{\Tr}{\mathrm{Tr}}

\newcommand{\sgn}{\mathrm{sgn}}

\begin{document}

\title{Spin Versus Charge Density Wave Order in Graphene-like Systems}
\author{Y.~Araki}
\affiliation{Department of Physics, The University of Tokyo, Tokyo 113-0033, Japan}
\affiliation{Department of Physics and Astronomy, University of British Columbia, \\ Vancouver, British Columbia, Canada V6T 1Z1}
\author{G.~W.~Semenoff}
\affiliation{Department of Physics and Astronomy, University of British Columbia, \\ Vancouver, British Columbia, Canada V6T 1Z1}

\begin{abstract}
 A variational technique is used to study  
 sublattice symmetry breaking by strong  on-site and nearest neighbor interactions in
graphene.  When interactions are strong enough
to break sublattice symmetry, and with relative strengths characteristic of graphene,
a charge density wave Mott insulator is favored over the spin 
density wave condensates. In the spin density wave condensate we find
that introduction of a staggered on-site energy (quasiparticle mass) leads to a splitting of the fermi velocities and mass
gaps of the quasiparticle spin states.  

\end{abstract}
\pacs{73.22.Pr, 71.10.Fd, 71.27.+a, 11.80.Fv} 
\maketitle

The possibility of gapping the spectrum of graphene,
either by explicit \cite{mindthegap} or spontaneous sublattice symmetry breaking \cite{Semenoff:2011jf},
is an important fundamental and practical problem \cite{ahcn}.
At the fundamental level, the question of spontaneous breaking of either exact or approximate chiral symmetry
emulates similar issues in quantum field theories such as quantum chromodynamics.
At the practical level, a small gap, particularly one which could be switched on and off
would be important for using graphene in electronics technology
as it could give a mechanism for controlling the flow of electrons.

In the absence of magnetic fields, the best
clean, suspended graphene is a semi-metal with no discernible energy gap.
Gap formation by spontaneous symmetry breaking, if it occurred,
would be driven by strong electron-electron interactions.
Numerical Monte Carlo computations and series expansions of the Hubbard model on a hexagonal lattice indicate
that a phase transition from a semi-metal to an anti-ferromagnetic, or spin density wave (SDW), Mott insulator \cite{sorella,pavia,stut}
(with perhaps other exotic phases in between)  will occur for relatively strong coupling, $U/t\sim $3-5,
where $U$ is the on-site Hubbard interaction and $t$ is the hopping parameter.
Estimates of these parameters for suspended graphene,
where an on-site Coulomb energy is $U\sim 10$eV and  $t=2.7$eV have a ratio in the same range,
raising the tantalizing idea that graphene is close to this critical point and some small modification 
which enhances the interaction could
induce a phase transition to a gapped state \cite{ahcn,herbut,drut,siogap,ahcn2}.
The simplest gapped states are spin density wave and charge density wave (CDW) Mott insulators, although
 more exotic phases have been discussed \cite{stut,herbut2,chamon,Raghu_2008,MacDonald_2012}.
There is also a possibility of breaking the sublattice symmetry explicitly
by depositing graphene on the appropriate substrate, such as boron nitride or silicon carbide \cite{mindthegap,siogap,bn}.
Even once it is broken explicitly, there can be phase transition between different patterns, for example CDW to SDW,
which can be of great interest. Moreover, the 
interplay between spontaneous and explicit symmetry breaking is an interesting problem which
has been discussed in recent literature \cite{Dillenschneider_2008,Araki_2011,Soriano_2011}.

\begin{figure}[tb]
\begin{center}
\includegraphics[width=7.0cm]{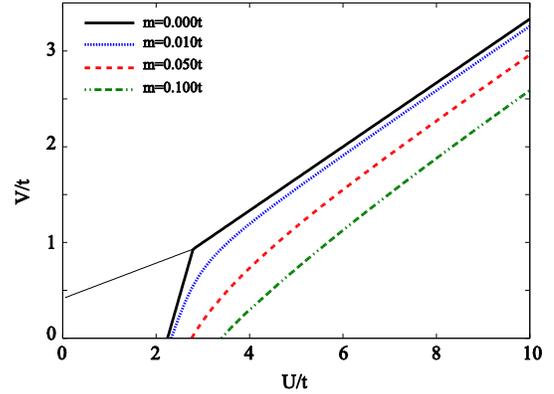}
\end{center}
\vspace{-0.5cm}
\caption{The phase diagram of the extended Hubbard model with  staggered potential  $m$.
The thick lines are phase boundaries between the SDW phase ($\sgn\Delta_\uparrow =-\sgn\Delta_\downarrow$) and the SM/CDW phase ($\sgn\Delta_\uparrow =\sgn\Delta_\downarrow$),
while the thin line for $m=0$ is the boundary between the SM and the CDW phases.
The SM phase does not appear when $m$ is finite.
As $m$ becomes larger, the SDW phase is suppressed.}
\label{fig:uv_m}
\end{figure}

In this Letter, we shall show that, even with explicit symmetry breaking, 
electron-electron interactions can change the character of the gap and the electron spectrum significantly.
For example, a candidate for the gapped phase is an antiferromagnetic SDW Mott insulator,
and it is indeed what is found in the Hubbard model at strong coupling \cite{sorella,pavia,stut}.
We shall show that, when next-to-nearest neighbor (NN) interactions are added to the Hubbard model,
with the  strength appropriate to graphene ($V\sim 10$eV),
a CDW state is favored over the antiferromagnetic SDW. A quantum phase transition between the two can be
driven by varying either the strength of the NN coupling or the amplitude of an explicit symmetry breaking
staggered potential.  
Our central result is the phase diagram in Fig.~\ref{fig:uv_m}.  The critical Hubbard coupling is underestimated by our technique, likely because magnon
fluctuations are not taken into account. In the absence of explicit symmetry breaking, the semi-metal
phase occurs in the trapezoid in the lower left-hand corner. 

We shall use a variational technique where we replace the full Hamiltonian $H$
by a solvable trial Hamiltonian $H_0$ which is optimized using  Jensen's inequality \cite{jensen},
\begin{equation} 
F \leq F_0 +\langle H-H_0 \rangle_0. \label{jensen}
\end{equation}
We shall adjust $H_0$ to minimize this upper bound on the free energy.
Here, $\langle {\mathcal O} \rangle_0=\Tr e^{-\beta H_0} {\mathcal O}/\Tr e^{-\beta H_0} $.

\begin{figure}[tb]
\begin{center}
\includegraphics[width=7.0cm]{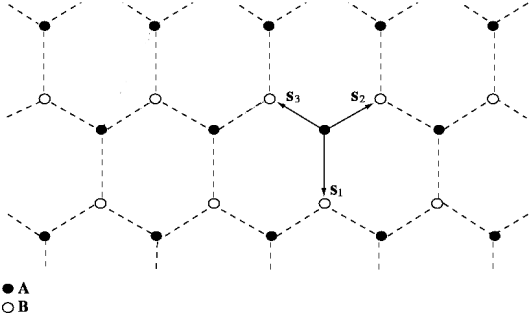}
\end{center}
\vspace{-0.5cm}
\caption{The hexagonal graphene lattice composed of sublattices A (black dots) and B (white dots)  connected
by the basis vectors ${\bf s}_i$.}
\label{sublattices}
\end{figure}

For the Hamiltonian of graphene, we begin with the tight-binding model with nearest-neighbor (NN) hopping.
\begin{equation}
H_T = -t\sum_{i,\sigma,\bfr \in A} \left[a_\sigma^\dag(\bfr)b_\sigma(\bfr+\bfs_i)+ b_\sigma^\dag(\bfr+\bfs_i)a_\sigma(\bfr)\right]. \label{ht}
\end{equation}
The hexagonal graphene lattice is depicted in Fig.~\ref{sublattices}.
It  contains two triangular sublattices, $A$ and $B$.
Creation and annihilation operators for electrons at sites $\bfr$ on sublattice A
are   $(a^\dag_{ \sigma}(\bfr) , a_{\sigma}(\bfr))$
and  $B$ are  $(b^\dag_{\sigma}(\bfr),b_{\sigma}(\bfr))$.
$\sigma=\uparrow,\downarrow$ is the spin index.
We shall add  a staggered on-site energy, $H_M$, which models explicit sublattice symmetry breaking
(which could arise by interaction with a substrate, for example, and gives the low energy graphene Dirac electron a mass gap \cite{Semenoff:1984dq}),  a Hubbard interaction $H_U$ and a NN interaction $H_V$, 
\begin{align}
H_M &= m\sum_{\bfr \in A} \left[b_\sigma^\dag(\bfr+\bfs_1)b_\sigma(\bfr+\bfs_1) -a_\sigma^\dag(\bfr)a_\sigma(\bfr) \right] \label{hm}\\
H_U&=  \frac{U}{2}\left[\sum_{\bfr\in A}\left(  a^\dag_{ \sigma}(\bfr)a_{\sigma}(\bfr)-1\right)^2 +\sum_{\bfr\in B}\left( b^\dag_{\sigma}(\bfr)b_{\sigma}(\bfr)-1\right)^2\right]\label{hu}\\
H_V&=V\sum_{\bfr\in A,i} \left[  a^\dag_{\sigma}(\bfr)a_{\sigma}(\bfr)-1\right] \left[ b^\dag_{\sigma'}(\bfr+\bfs_i)b_{\sigma'}(\bfr+\bfs_i)-1\right] \label{hv}
\end{align}
where terms such as $ a^\dag_{ \sigma}(\bfr)a_{\sigma}(\bfr)$ are summed over spins.
An important symmetry of graphene
which is to a good approximation  visible in  angle-resolved photoemission spectroscopy (ARPES) measurements \cite{arpes}
is particle-hole symmetry.
Here, we have written a model Hamiltonian $H=H_T+H_M+H_U+H_V$ which has exact particle-hole symmetry.
We will also restrict the variational Ans\"atz to have this symmetry.
The explicit particle-hole transformation is
$
a^\dag_{\sigma}(\bfr) , a_{\sigma}(\bfr),b^\dag_{\sigma}(\bfr),b_{\sigma}(\bfr)$ $ \to$  $ a_{\sigma}(\bfr) , a^\dag_{\sigma}(\bfr),-b_{\sigma}(\bfr),-b^\dag_{\sigma}(\bfr)$.
The terms in the Hamiltonian $H_T,H_U,H_V,H_M$ are invariant.

To write down the trial Hamiltonian $H_0$,
it is convenient to Fourier transform to momentum space where
\begin{equation}
H_0 = \sum_{\bfk,\sigma} (a_\sigma^\dag(\bfk),b_\sigma^\dag(\bfk)) \left(
\begin{array}{cc}
\Delta_\sigma(\bfk) & h_\sigma(\bfk) \\
h_\sigma^*(\bfk) & -\Delta_\sigma(\bfk)
\end{array}
\right)
\left(
\begin{array}{c}
a_\sigma(\bfk) \\ b_\sigma(\bfk)
\end{array}
\right),
\end{equation}
where $\bfk$ is a wave-vector in the Brillouin zone of the triangular lattice,
and, for example
\begin{equation}
a(\bfk)=\sum_{\bfr\in A} \frac{e^{i\bfk\cdot \bfr}}{\sqrt{\Omega}} a_{\sigma}(\bfr)
~,~ a_{\sigma}(\bfr)=\int d\bfk\frac{ e^{-i\bfk\cdot \bfr}}{\sqrt{ \Omega}}a_\sigma(\bfk)
\end{equation}
with $ {\Omega}$ the volume of the Brillouin zone.
Here we assume that the different matrix elements in the Hamiltonian can be simultaneously diagonalized in spin.
This is not the most general possible Ans\"atz,  which would have a more complicated spin dependence.
We have assumed translation invariance on the triangular sublattices.
If we set $\Delta_\sigma(\bfk)=0$ and
$h_\sigma(\bfk)=\sum e^{i\bfk\cdot{\bf s}_i}\equiv\Phi(\bfk)$,
$H_0$ becomes identical to the tight-binding model Hamiltonian $H_T$.
We have fixed the diagonal parts of $H_0$ so that it has particle-hole symmetry.
Aside from particle-hole symmetry,  $H_T,H_U,H_V$ also have sublattice symmetry
-- where we simply interchange the sublattice excitations
$
a^\dag_\sigma(\bfk),a_\sigma(\bfk), b^\dag_{\sigma}(\bfk),b_\sigma(\bfk)$ $\to$
 $b^\dag_\sigma(-\bfk),b_\sigma(-\bfk), a^\dag_{\sigma}(-\bfk),a_\sigma(-\bfk)$.
This symmetry is broken by $H_M$, which flips sign under the transformation.
The trial Hamiltonian has this symmetry only when $\Delta_\sigma=0$.
Hermiticity requires that $h_\sigma^*(-\bfk)=h_\sigma(\bfk)$ and $\Delta_\sigma(\bfk)=\Delta_\sigma(-\bfk)=\mathrm{real}$.

The spectrum and the eigenstates of $H_0$ are easy to find:
The eigenvalues of the single-particle Hamiltonian are
$E_{\sigma,\pm}(\bfk)\equiv \pm E_\sigma(\bfk) =\pm\sqrt{\Delta_\sigma(\bfk)^2+|h_\sigma(\bfk)|^2}$.
With a change of variables into polar coordinate $h=E\cos\theta e^{i\phi}~,~\Delta=E\sin\theta~ \; (-\pi/2 \leq \theta \leq \pi/2,~-\pi < \phi \leq \pi)$,
$H_0$ is diagonalized by the canonical transformation
\begin{eqnarray}
a &=& \frac{1}{\sqrt{2(1+\sin\theta)}} \left[(1+\sin\theta)\psi_+ -\cos\theta e^{i\phi} \psi_-\right] \\
b &=& \frac{1}{\sqrt{2(1+\sin\theta)}} \left[\cos\theta e^{-i\phi}\psi_+ + (1+\sin\theta)\psi_-\right] ,
\end{eqnarray}
where we have suppressed $\bfk,\sigma$ labels,
and $(\psi_+^\dag,\psi_+)$ and $(\psi_-^\dag,\psi_-)$ are creation and annihilation operators for
electrons in energy states $+E_\sigma(\bfk)$ and $-E_\sigma(\bfk)$, respectively.
With this transformation, the correlation functions are diagonal in momentum and spin space,
\begin{eqnarray}
\langle a^\dag_\sigma (\bfk) a_{\sigma}(\bfk)\rangle_0 &=& \tfrac{1}{2}\left[1-\sin\theta_\sigma(\bfk)\tanh\tfrac{\beta}{2}E_\sigma(\bfk)\right] \\
\langle b^\dag_\sigma (\bfk) b_{\sigma}(\bfk)\rangle_0 &=& \tfrac{1}{2}\left[1+\sin\theta_\sigma(\bfk)\tanh\tfrac{\beta}{2}E_\sigma(\bfk) \right] \\
\langle b^\dag_\sigma (\bfk) a_{\sigma}(\bfk)\rangle_0 &=& -\tfrac{1}{2} \cos\theta_\sigma(\bfk) e^{i\phi_\sigma(\bfk)} \tanh\tfrac{\beta E_\sigma(\bfk)}{2}.
\end{eqnarray}
All the others can be obtained from these by simple algebra.
All expectation values of operators factor into bilinears such as these.
Then, the free energy per unit volume is the sum of the following five contributions,
which come from $F_0-\langle H_0\rangle_0$ and
 the expectation values of $H_T, H_M, H_U, H_V$, respectively:
\begin{align}
& \epsilon_0=  \int\frac{d\bfk}{\Omega}\sum_{\sigma}\left[ E_\sigma(\bfk) \tanh\tfrac{\beta E_\sigma(\bfk)}{2} -\tfrac{2}{\beta}\ln\left[2\cosh\tfrac{\beta E_\sigma(\bfk)}{2}\right]\right] \label{1} \\
& \epsilon_T=\frac{t}{2}\int \frac{d\bfk}{\Omega}\sum_\sigma \cos\theta_\sigma(\bfk)e^{i\phi_\sigma(\bfk)}\Phi(\bfk)\tanh\tfrac{\beta}{2}E_\sigma(\bfk) +{\rm c.c.} \label{2} \\
& \epsilon_M=  m\int\frac{d\bfk}{\Omega}\sum_\sigma  \sin\theta_\sigma(\bfk)\tanh\tfrac{\beta}{2}E_\sigma(\bfk) \label{3} \\
& \epsilon_U= \frac{U}{4} \left[\sum_\sigma\int\frac{d\bfk}{\Omega}  \sin\theta_\sigma(\bfk)\tanh\tfrac{\beta}{2}E_\sigma(\bfk)
\right]^2 \nonumber \\
& \quad \quad \quad -\frac{U}{4}\sum_\sigma \left[
\int\frac{d\bfk}{\Omega}  \sin\theta_\sigma(\bfk)\tanh\tfrac{\beta}{2}E_\sigma(\bfk)
\right]^2 \label{4} \\
& \epsilon_V= -\frac{3V}{4}\left[\sum_\sigma\int\frac{d\bfk}{\Omega}  \sin\theta_\sigma(\bfk)\tanh\tfrac{\beta}{2}E_\sigma(\bfk)
\right]^2  \nonumber \\
& -\frac{V}{12}\sum_\sigma\left|\int\frac{d\bfk}{\Omega} \cos\theta_\sigma(\bfk)e^{i\phi_\sigma(\bfk)}\Phi(\bfk)\tanh\tfrac{\beta}{2}E_\sigma(\bfk)\right|^2 \label{5}
\end{align}
First, consider the equation obtained from varying $\phi_\sigma(\bfk)$:
\begin{equation}
0 = Z_\sigma e^{i\phi_\sigma(\bfk)}\Phi(\bfk) -Z^*_\sigma e^{-i\phi_\sigma(\bfk)}\Phi^*(\bfk),
\end{equation}
where the factor $Z_\sigma$ is defined by
\begin{equation}
Z_\sigma = 1- \frac{V}{6t} \int\frac{d\bfk}{\Omega}  \cos\theta_\sigma(\bfk)e^{-i\phi_\sigma(\bfk)}\Phi(\bfk)\tanh\frac{\beta }{2}E_\sigma(\bfk).
\end{equation}
The solution of this equation which minimizes the energy is $\phi_\sigma(\bfk) =-\arg \Phi(\bfk)+\pi$.
Thus, everywhere in Eqs.~(\ref{1})-(\ref{5}), $e^{i\phi}\Phi$ can be replaced by $-|\Phi|$.
The equation obtained by varying $E_\sigma(\bfk)$ and  $\theta_\sigma(\bfk)$ are
\begin{align}
E_\sigma(\bfk) & = \cos\theta_\sigma(\bfk)  Z_\sigma t|\Phi(\bfk)| \nonumber \\
 &+ \left[\tfrac{3V}{2}C_\sigma-m+\tfrac{3V-U}{2}C_{\bar{\sigma}} \right]\sin\theta_\sigma(\bfk), \label{e} \\
Z_\sigma t|\Phi(\bfk)|&\tan\theta_\sigma(\bfk)=  \tfrac{3V}{2} C_\sigma -m + \tfrac{3V-U}{2}C_{\bar\sigma},
\end{align}
where
\begin{align}
C_\sigma&=\int \frac{d\bfk}{\Omega}\sin\theta_\sigma(\bfk)\tanh\frac{\beta }{2}E_\sigma(\bfk) \label{c} \\
Z_\sigma &= 1+ \frac{V}{6t}\int \frac{d\bfk'}{\Omega} \cos\theta_\sigma(\bfk') |\Phi(\bfk)|\tanh\frac{\beta }{2}E_\sigma(\bfk). \label{z}
\end{align}
The solution reads
\begin{align}
&E_\sigma(\bfk)=\sqrt{ Z_\sigma^2 t^2|\Phi(\bfk)|^2+\left[\tfrac{3V}{2}C_\sigma-m+\tfrac{3V-U}{2}C_{\bar{\sigma}} \right]^2} \label{e1} \\
&\sin\theta_\sigma(\bfk)=\left[\tfrac{3V}{2}C_\sigma-m+\tfrac{3V-U}{2}C_{\bar{\sigma}} \right]/E_\sigma(\bfk) \label{sin} \\
&\cos\theta_\sigma(\bfk)= Z_\sigma t|\Phi(\bfk)| / E_\sigma(\bfk), \label{cos}
\end{align}
The four constants $C_\sigma$ and $Z_\sigma$ must be determined self-consistently.
  $Z_\sigma$  corrects the fermi velocity and
$C_\sigma$ and $m$ gap the spectrum.
If $m$ were zero, but $C_\sigma$ nonzero, the sublattice symmetry would be spontaneously broken.
The nonzero temperature is important for deriving the variational equations, however,
to study the low temperature limit, we will set it to zero.

Since, from Eq.~(\ref{4}), $\epsilon_U=\frac{U}{2}C_\uparrow C_\downarrow$,
the Hubbard interaction favors a spin density wave (SDW)
where $C_\uparrow$ and $C_\downarrow$ are nonzero and have opposite signs.
From Eq.~(\ref{5}), $\epsilon_V= -\frac{3V}{4}(C_\uparrow+C_\downarrow)^2-3V[(Z_\uparrow-1)^2+(Z_\downarrow-1)^2] $.
The NN interaction favors a charge density wave (CDW) where $C_\uparrow$ and $C_\downarrow$ are nonzero and
have the same sign.
The competition of these two phases is seen in the numerical solutions of the self-consistent equations, Eqs.~(\ref{e})-(\ref{cos}).
The phase diagram is shown in Fig.~\ref{fig:uv_m}.
When $m=0$, there are three phases:
a semi-metal (SM) for $U,V \lesssim t$, SDW for $U \gtrsim V,m$, and CDW for $V,m \gtrsim U$.
$H_M$ is a source for CDW. When it is finite, there is no SM phase. The SDW phase is suppressed as $m$ increases, while the CDW phase is enhanced.
 
\begin{figure}[tb]
\begin{center}
\begin{tabular}{cc}
\includegraphics[width=4.1cm]{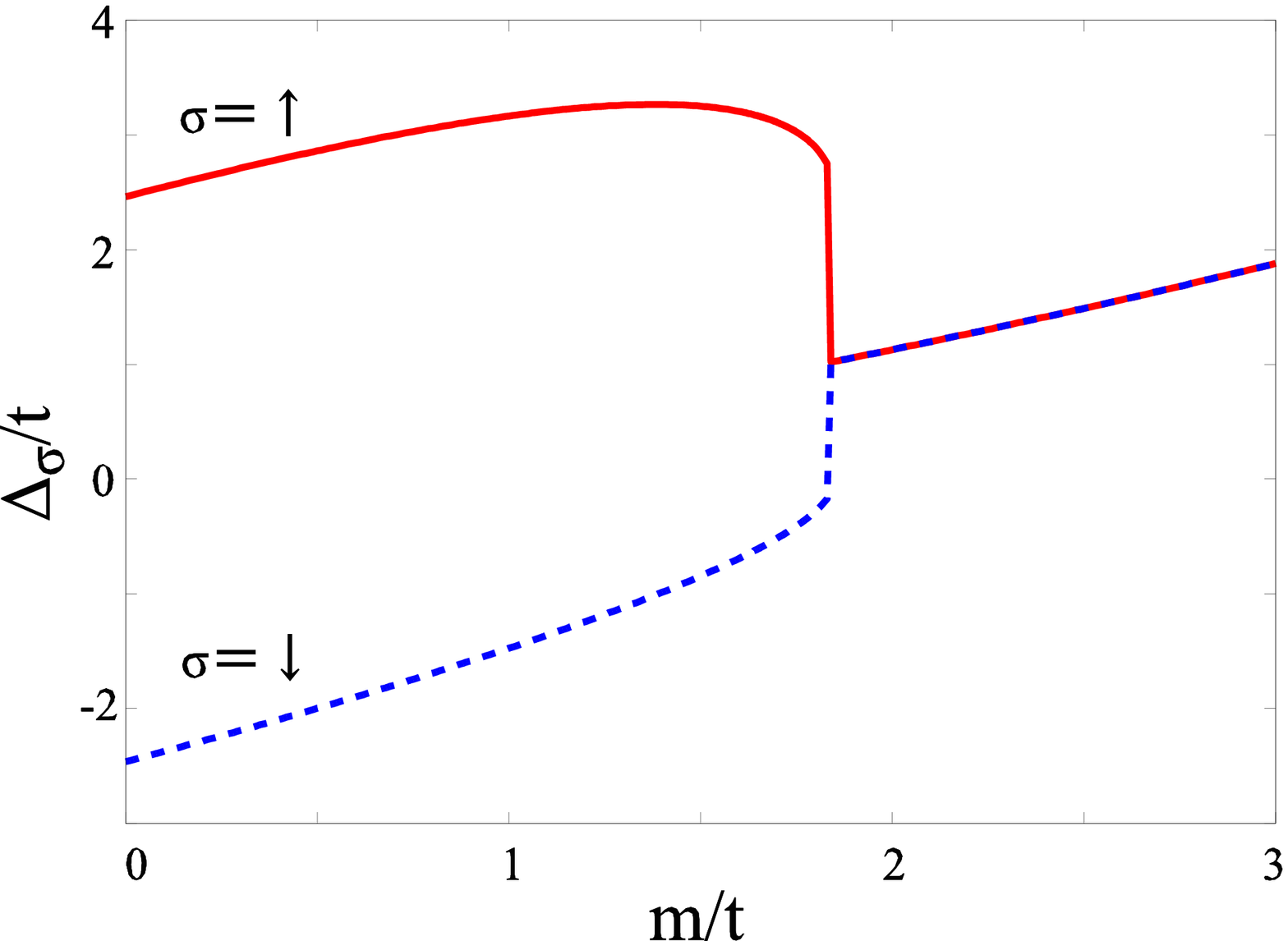} & \includegraphics[width=4.1cm]{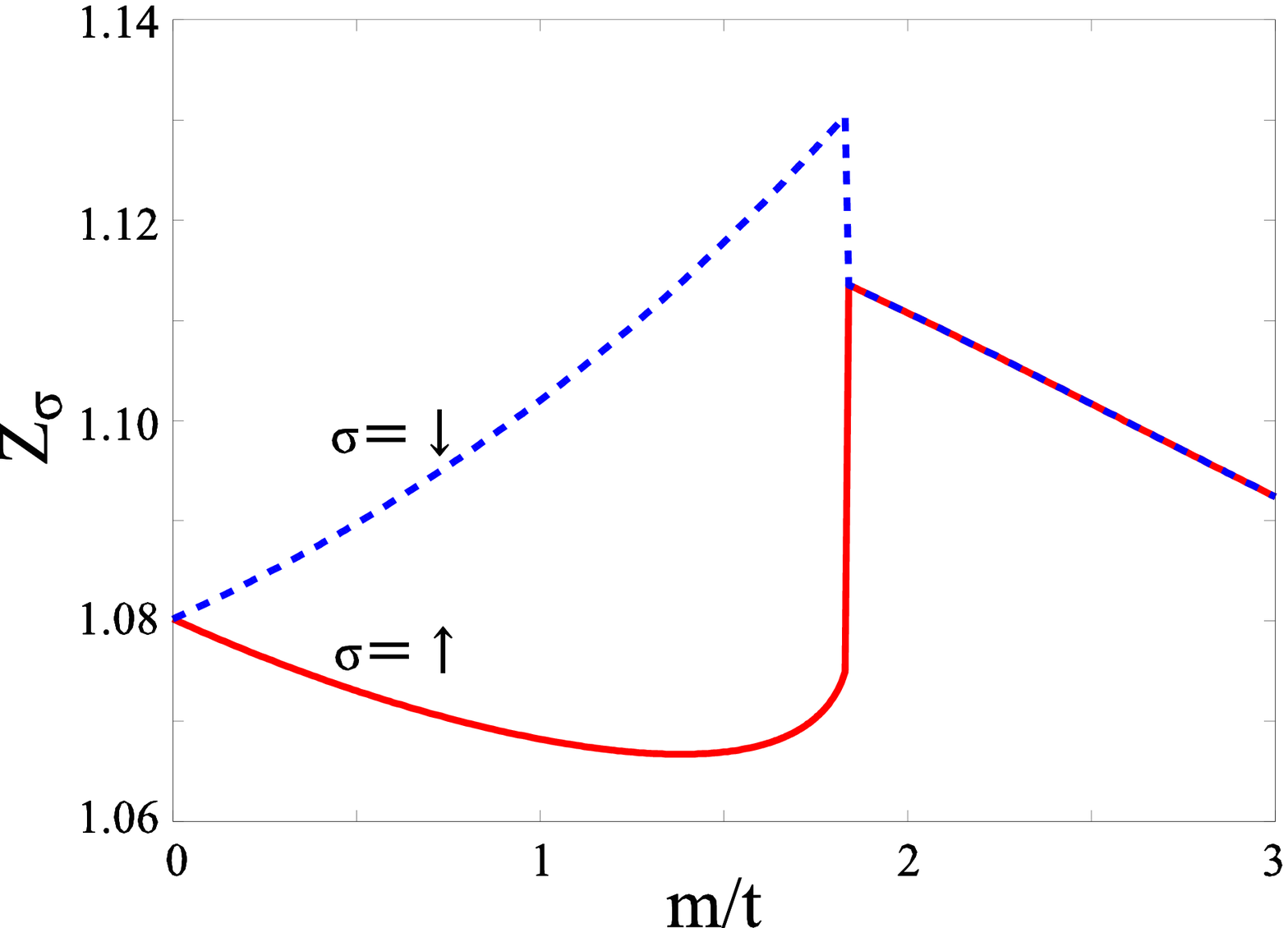}
\end{tabular}
\end{center}
\vspace{-0.5cm}
\caption{The behavior of the density wave amplitude $\Delta_\sigma$ (left) and the velocity renormalization factor $Z_\sigma$ (right)
as functions of the external mass $m$, where the on-site interaction $U=6.0t$ and the NN interaction $V=0.5t$ are fixed.
The system shows the SDW phase for $m<1.8t$, while it reveals the CDW phase for $m>1.8t$.
The fermi velocity of the up spin and that of the down spin differ (i.e. $Z_\uparrow \neq Z_\downarrow$) in the SDW phase, unless $m=0$.}
\label{fig:uvhm}
\end{figure}

Now we shall investigate the quantitative behavior of $\Delta_\sigma$ and $Z_\sigma$,
by varying one parameter out of $U$, $V$, $m$ while holding the others fixed.
We begin with $U=6.0t,\;V=0.5t,\;m=0$, where the system is in the SDW phase,
and we increase $m$.
As shown in the left panel of Fig.~\ref{fig:uvhm},
both $\Delta_\uparrow$ and $\Delta_\downarrow$ increase as a function of $m$ as long as $m$ is sufficiently small.
It should be noted that $|\Delta_\uparrow|$ and $|\Delta_\downarrow|$ take different values in this region unless $m=0$,
which means that the quasiparticle gap for up spin and that for down spin are different.
Such a discrepancy of $|\Delta_\sigma|$ also causes the discrepancy of the   factor $Z_\sigma$ through Eq.(\ref{z}),
as shown in the right panel of Fig.\ref{fig:uvhm}.
The difference between $Z_\uparrow$ and $Z_\downarrow$ increases as a function of $m$
towards its maximum value $Z_\downarrow-Z_\uparrow=0.05$ at $m=1.7t$,
then drastically drops towards zero at the critical value $m_C=1.8t$.
Since $\Delta_\uparrow=\Delta_\downarrow$ in the CDW region,
quasiparticles with up spin and those with down spin obtain the same Fermi velocity above $m_C$.

\begin{figure}[tb]
\begin{center}
\begin{tabular}{cc}
\includegraphics[width=4.1cm]{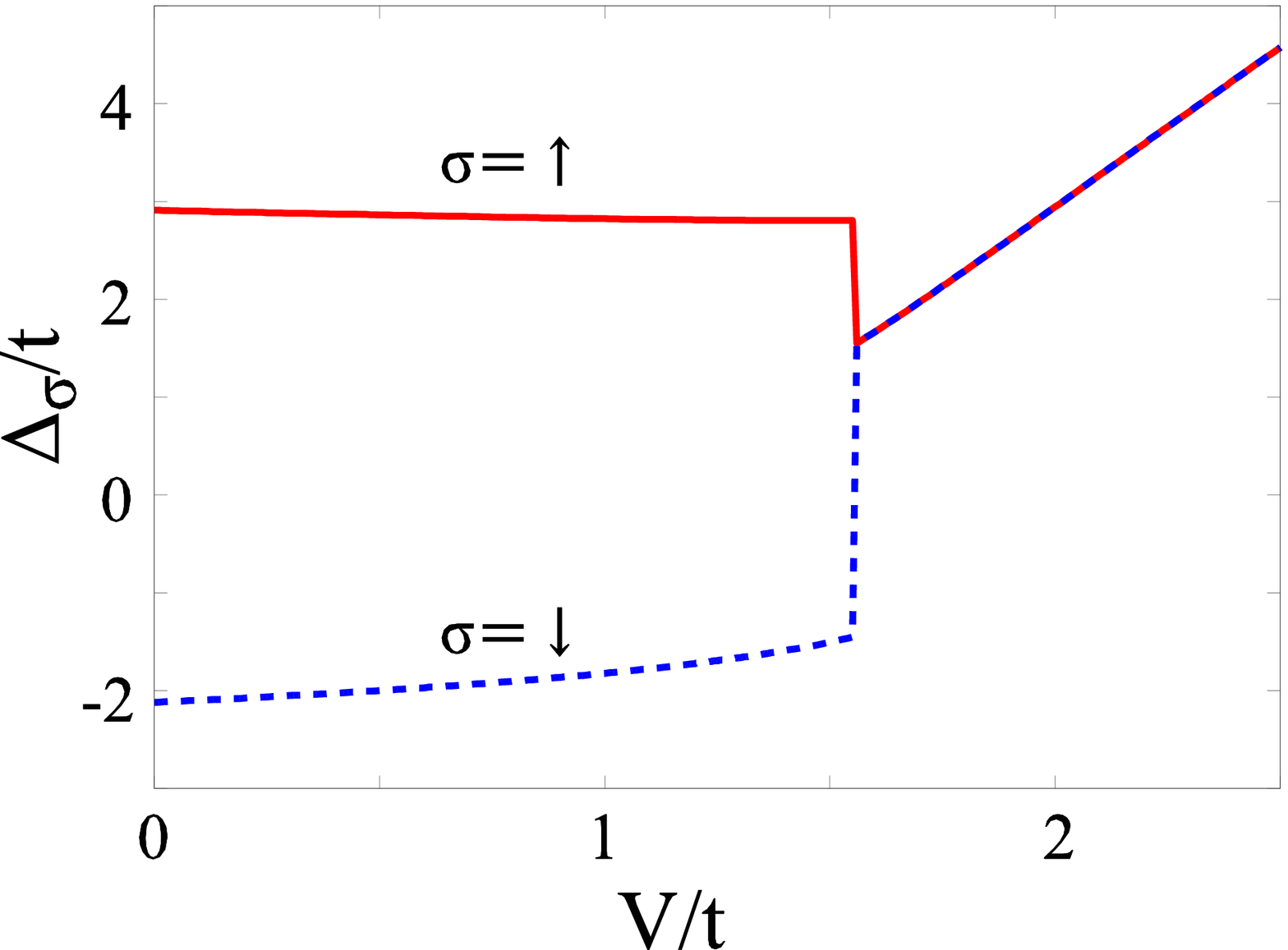} & \includegraphics[width=4.1cm]{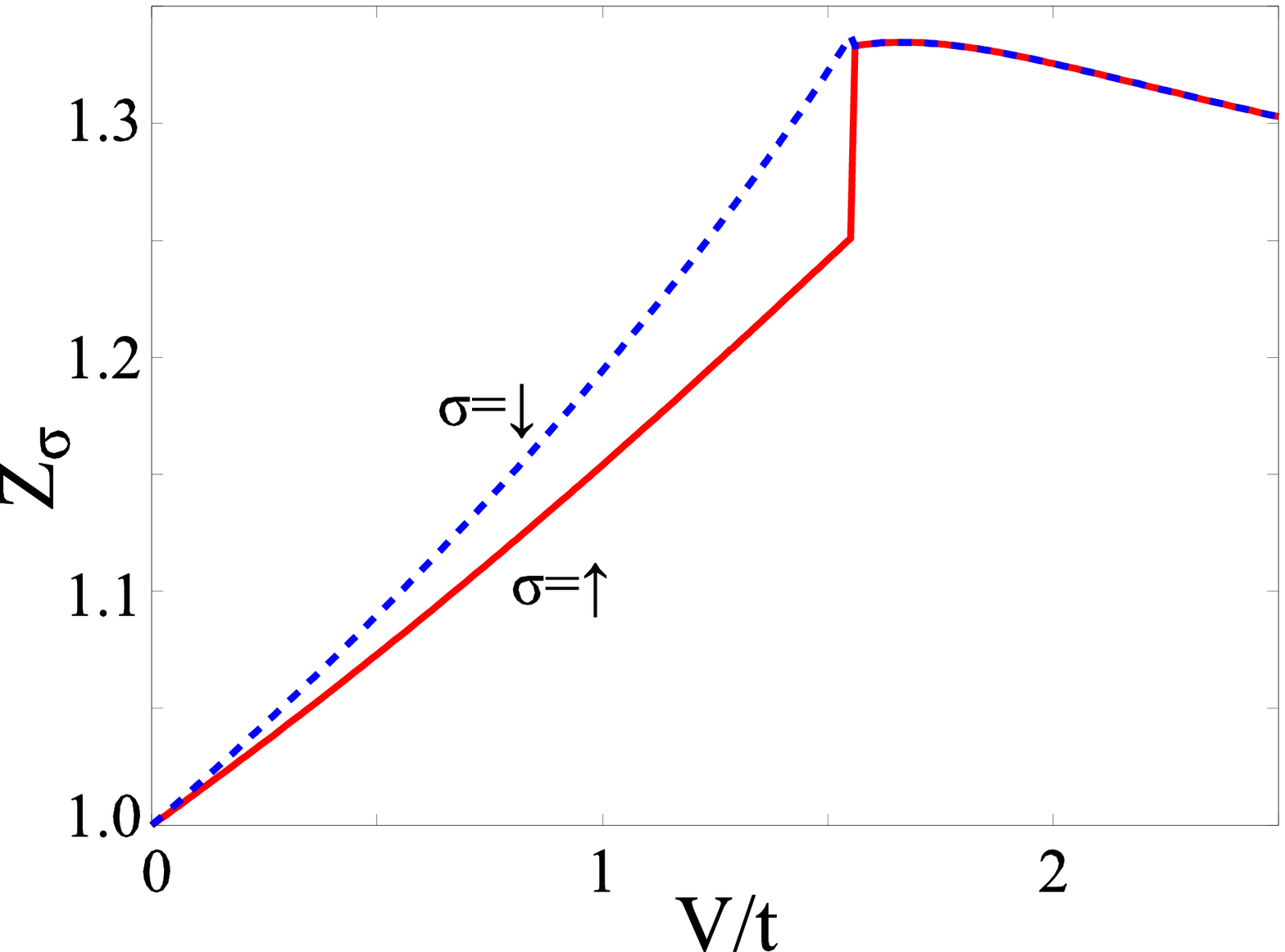}
\end{tabular}
\end{center}
\vspace{-0.5cm}
\caption{The behavior of the density wave amplitude $\Delta_\sigma$ (left) and the velocity renormalization factor $Z_\sigma$ (right)
as functions of the NN interaction $V$, where the on-site interaction $U=6.0t$ and the external mass $m=0.5t$ are fixed.
The system shows the SDW phase for $V<1.5t$, while it reveals the CDW phase for $V>1.5t$.
The fermi velocity of the up spin and that of the down spin differ (i.e. $Z_\uparrow \neq Z_\downarrow$) in the SDW phase, unless $V=0$.}
\label{fig:umhv}
\end{figure}

Next we vary the NN interaction $V$,
where the on-site interaction $U=6.0t$ and the mass $m=0.5t$ are fixed.
The SDW amplitude is suppressed as $V$ increases,
and a phase transition  to the CDW phase occurs at $V_C=1.5t$,
as shown in the left panel of Fig.\ref{fig:umhv}.
Due to the finite external mass, $m$,
there is a discrepancy between $|\Delta_\uparrow|$ and $|\Delta_\downarrow|$ in the SDW phase,
which leads to the discrepancy between $Z_\uparrow$ and $Z_\downarrow$,
as shown in the right panel of Fig.\ref{fig:umhv}.
Since $Z_\sigma-1$ is proportional to $V$,
$Z_\uparrow$ and $Z_\downarrow$ take the identical value (unity) at $V=0$,
even though the quasiparticle gap amplitudes are different.
$Z_\downarrow-Z_\uparrow$ reaches its maximum value $0.09$ just below the critical value $V_C$.

In conclusion, we note that, in the continuum limit of graphene, the CDW and SDW
condensates are indistinguishable as they are related to each other by a transformation in the emergent U(4) symmetry.  
We have found that they are indeed distinguished by lattice scale physics which can have an important effect. 
We have shown
that the short ranged interactions of relative strengths approximating graphene favor the CDW state.  This is basically due to the fact that the on-site energy  is anomalously small compared
to the NN potential energy.   Explicit symmetry breaking, which can
be present in some cases enhances this effect.  The conclusion that lattice scale physics can drive
a phase transition is surprising.  It is likely that the naive continuum  Coulomb interaction is good for the semi-metal phase, however when density wave order sets in, it is driven by otherwise irrelevant four-Fermion interactions which can have nontrivial strong coupling fixed points.  This point of view is supported by renormalization group analyses of the continuum theory \cite{rg}. Another anomalous effect of explicit symmetry breaking is the splitting of the Fermi velocities of the spin up and
spin down electrons in the SDW phase. That splitting goes to zero if $m$ goes to zero.  It increases as a function of the NN interaction strength,
and it reaches about $10\%$ of the Fermi velocity.
Such a discrepancy might be detected by ARPES measurements,
and it may influence transport properties of the system.

Some of our results are similar to a self-consistent mean field theory. 
We point out that the variational technique is more general in that it contains a wave-function renormalization,
which is normally absent in mean field approach.
It is also readily applicable to a much wider array of potentials,
and we believe that our exposition of the technique here could be used as a starting point
for more general analyses of graphene-like systems. 
We have focused on the SDW and CDW patterns,
but the honeycomb lattice can have a richer array of symmetry breaking patterns,
such as the Kekul\'e distortion which is expected to become relevant when the 
next-to-NN interaction is taken into account. 
The interplay of ordering patterns including those phases,
induced either spontaneously or explicitly,
remains an open question.

\

\begin{acknowledgments}
Y.~A.~ is supported by Grant-in-Aid for Japan Society for the Promotion of Science (DC1, No.22.8037).
G.W.S.~is supported by NSERC of Canada.
\end{acknowledgments}


\end{document}